\documentstyle[12pt,preprint,aps,prb,eqsecnum]{revtex}
\def\nm{\nonumber}
\def\beqa{\begin{eqnarray}}
\def\beq{\begin{equation}}
\def\F{{\cal{F}}}

\def\eeqa{\end{eqnarray}}
\def\eeq{\end{equation}}
\def\lab{\label}    
\def\pa{\partial}
\def\l{\Lambda}

\begin{document}

\begin{titlepage}
\thispagestyle{plain}
\pagenumbering{arabic}
\begin{center}
{\Large \bf 
Dual WDVV Equations in }
\end{center}
\vspace{-7.0mm}
\begin{center}
{\Large \bf $N=2$ Supersymmetric Yang-Mills Theory}
\end{center} 
\vspace{-7.0mm}
\lineskip .80em
\vskip 4em
\normalsize
\begin{center}
{\large Y\H uji Ohta}
\end{center}
\vskip 1.5em
\begin{center}
{\em Research Institute for Mathematical Sciences }
\end{center}
\vspace{-11.0mm}
\begin{center}
{\em Kyoto University}
\end{center}
\vspace{-11.0mm}
\begin{center}
{\em Sakyoku, Kyoto 606, Japan.}
\end{center}
\begin{abstract}
This paper studies the dual form of 
Witten-Dijkgraaf-Verlinde-Verlinde (WDVV) 
equations in $N=2$ supersymmetric 
Yang-Mills theory by applying a duality 
transformation to WDVV equations. The dual WDVV equations called 
in this paper are non-linear differential equations satisfied by 
dual prepotential and are found to have the same form with the original 
WDVV equations. However, in contrast with the case of 
weak coupling calculus, the perturbative part of dual 
prepotential itself does not satisfy the dual WDVV equations. 
Nevertheless, it is possible to show that the non-perturbative part of 
dual prepotential can be determined from dual WDVV equations, 
provided the perturbative part is given. As an example, 
the SU(4) case is presented. The non-perturbative dual prepotential 
derived in this way is consistent to the dual prepotential obtained 
by D'Hoker and Phong. 
\end{abstract}
\end{titlepage}

\begin{center}
\section{Introduction}
\end{center}

We have had a rich harvest since the seed of approach using a Riemann 
surface for the low energy effective description of $N=2$ supersymmetric 
Yang-Mills theory was sowed by Seiberg and Witten. \cite{SW1,SW2} For 
example, instanton effect \cite{Sei} for prepotentials obtained by using 
a Riemann surface \cite{KLT} showed a good agreement to the prediction of 
instanton calculus, \cite{FP,DKM,DKM2} integrable structure behind 
Seiberg-Witten solutions was discussed in 
terms of Whitham theory, \cite{GKMMM} and the approach taken by 
Seiberg and Witten was extended to the case of higher dimensional 
gauge theories. \cite{Nek} Of course, there are a lot of other interesting 
developments, but the best way to understand uniformly all these aspects 
of $N=2$ supersymmetric gauge theories may be 
encoded in the language of Witten-Dijkgraaf-Verlinde-Verlinde (WDVV) 
equations \cite{Wit,DW,DVV,Dub,MMM,MMM2,MMM3,IY2} because they can widely 
cover various aspects of prepotentials. 

In general, WDVV equations hold in various dimensional gauge 
theories \cite{MMM2,BMMM} and (in the case not including massive 
matter hypermultiplets) they are of the form
	\beq
	(\F_{i} )(\F_{k})^{-1}(\F_{j})=(\F_{j})(\F_{k})^{-1}(\F_{i})
	,\lab{wdvv}
	\eeq
where $\F$ is the prepotential, 
$(\F_i )_{jk}:= \pa^3 \F/\pa a_i \pa a_j \pa a_k$ are 
matrix notations, $a_i$ are regarded as periods of Seiberg-Witten 
differential and the indices run from 1 to the rank of the 
gauge group. (\ref{wdvv}) was extensively investigated at perturbative 
level, \cite{MMM,MMM2,MMM3,IY2} but not so much about the 
non-perturbative effect obtainable from (\ref{wdvv}) are known. \cite{O2} 

On the other hand, the study of strong coupling region in view of 
WDVV equations are not found in literatures. According to the 
electro-magnetic duality of Seiberg and Witten, \cite{SW1,SW2} 
in the strong coupling region where charged particles become massless, 
the role of periods $a_i$ and their magnetic duals $a_{D_j}:=\pa \F /a_j$ 
are exchanged. If this duality is applied to third-order derivatives of 
prepotential, we will obtain non-linear equations like (\ref{wdvv}) written 
in terms of dual periods. The equations obtained in this way will make it 
possible to derive the strong coupling prepotentials in the standpoint of 
WDVV equations. The aim of the paper is to study such non-linear 
equations. 

This paper is organized as follows. In Sec. II, we apply the duality 
transformation to third-order derivatives of prepotential and by using 
(\ref{wdvv}) we derive non-linear equations for dual prepotential. These 
equations are found to have the same form with (\ref{wdvv}), so we call 
them as dual WDVV equations throughout the paper. In Sec. III, 
we consider the relation among dual prepotentials and the dual WDVV 
equations. In particular, we firstly show in SU($r+1$) gauge theory that 
the dual perturbative prepotential \cite{DS,DP} 
do not satisfy dual WDVV equations. Of course, in order to determine 
non-perturbative part, the perturbative part must be required, so we 
provide it as input data. However, as the general case is slightly 
intractable, we present the calculation of non-perturbative part of dual 
prepotential in SU(4) gauge theory as an example. We can find that 
the non-perturbative dual prepotential which is consistent to that 
found by D'Hoker and 
Phong \cite{DP} is available from dual WDVV equations. 
Sec. IV is a brief summary. 

\begin{center}
\section{The dual WDVV equations}
\end{center}

In this section, we prove the existence of dual form of WDVV equations 
for all known models with WDVV equations (\ref{wdvv}). Our method here 
is based on the action of duality transformation for prepotential. 

To begin with, let us consider how the third-order derivatives of 
prepotential transform under the electro-magnetic duality. In general, it 
is well-known that in the case of 
rank $r$ gauge group the full electro-magnetic duality group is a 
subgroup of 
Sp($2r,\mbox{\boldmath$Z$}$), \cite{KLYT,APS,HO,DS1,Han,BL} 
and the generator
	\beq
	S:=\left(\begin{array}{rcc}
	0 & &I \\
	-I & &0
	\end{array}\right)
	,\lab{sd}
	\eeq
where $I$ and $0$ are the unit matrix and zero matrix of size $r\times r$, 
respectively, induces the exchange of periods and their duals and therefore 
the inversion of the effective coupling constant. Note that in order to 
see a strong coupling behavior it is enough to take into account of only 
(\ref{sd}) and is not necessary to consider all actions of duality group.

In fact, the periods transform under (\ref{sd}) as 
	\beq
	\left(\begin{array}{c}
	a_{D_i}\\
	a_j 
	\end{array}\right)\longrightarrow  
	\left(\begin{array}{c}
	\widetilde{a}_{D_i} \\
	\widetilde{a}_j 
	\end{array}\right):=S \left(\begin{array}{c}
	a_{D_i}\\
	a_j 
	\end{array}\right)
	.\eeq 
Then the effective coupling constants transform as 
	\beq
	\tau_{ij}:= \frac{\pa a_{D_i}}{\pa a_j}=\frac{\pa^2 \F}{\pa a_i \pa a_j} 
	\longrightarrow 
	\tau_{D_{ij}}:=\frac{\pa \widetilde{a}_{D_i}}{\pa 
	\widetilde{a}_j}=-\frac{1}{\displaystyle \frac{\pa a_{D_j}}
	{\pa a_i}}=-\frac{1}{\tau_{ij}} 
	,\eeq
where $\tau_{D_{ij}}$ are the dual effective coupling constants. 
From this transformation property, it is immediate to see that 
	\beq
	\frac{\pa \tau_{D_{ij}}}{\pa \widetilde{a}_k}=
	\sum_{l=1}^{r}\frac{\pa }{\pa a_l}
	\left(-\frac{1}{\tau_{ij}}\right)\frac{\pa a_l}{\pa \widetilde{a}_k}
	=\sum_{l=1}^{r} 
	\frac{(\F_i )_{jl}}{\tau_{ij}^2}\tau_{D_{lk}}
	.\lab{www}
	\eeq
Note that in (\ref{www}) the repeated indices $i$ and $j$ are {\em not} 
summed. 

Now, suppose that $\widetilde{a}_{D_i}$ are given by differentiations of 
some function $\F_D$ (the dual prepotential) 
	\beq
	\widetilde{a}_{D_i}:=\frac{\pa \F_D}{\pa \widetilde{a}_i}=
	-\frac{\pa \F_D}{\pa a_{D_i}}
	.\eeq
Then the relation (\ref{www}) can be rewritten as  
	\beq
	(\F_{D_i})_{jk}=\sum_{l=1}^r (\F_i )_{jl} (\tau_D )_{lk}
	,\lab{matr}
	\eeq
where 
	\beq
	(\F_{D_i})_{jk}:=\frac{\pa^3 \F_D}{\pa \widetilde{a}_i \pa 
	\widetilde{a}_j \pa \widetilde{a}_k}=-
	\frac{\pa^3 \F_D}{\pa a_{D_i}\pa a_{D_j} \pa a_{D_k}}
	\lab{defFD}
	\eeq
and $(\tau_D )_{lk}:=\tau_{D_{lk}}$ are matrix notations and the overall 
factor is ignored because it is not necessary in the following 
discussions. Note that the right hand side of (\ref{matr}) is simply a 
matrix multiplication. 

With the aid of (\ref{matr}), it is easy to see that 
	\beq
	(\F_{D_i})_{pl}(\F_{D_k})_{lm}^{-1}(\F_{D_j})_{mn}=
	(\F_i )_{pl}(\F_k )_{lm}^{-1}(\F_j )_{mq}(\tau_D )_{qn}
	,\lab{228}
	\eeq
where $(\F_{D_k})_{lm}^{-1}$ mean the $(l,m)$ components 
of $(\F_{D_k})^{-1}$ and 
we have ignored the determinants of $(\F_k)$ and $(\tau_D)$ 
arising from $(\F_{D_k})^{-1}$ because they can be summarized into 
an overall factor. Thus from (\ref{228}) and (\ref{wdvv}), 
we can obtain the dual form of WDVV equations
	\beq
	(\F_{D_i} )(\F_{D_k})^{-1}(\F_{D_j})=(\F_{D_j})(\F_{D_k})^{-1}
	(\F_{D_i})
	.\lab{duawdvv}
	\eeq
Since (\ref{duawdvv}) is written by dual variables, we often refer 
(\ref{duawdvv}) as dual WDVV equations throughout the paper. 

From our construction, it would be obvious that there also exist dual 
WDVV equations if WDVV equations (\ref{wdvv}) hold.

\begin{center}
\section{Dual prepotential available from dual WDVV equations}
\end{center}

The dual WDVV equations (\ref{duawdvv}) take the same form 
with (\ref{wdvv}), but the study of dual prepotential in strong 
coupling region from a standpoint of (\ref{duawdvv}) is slightly 
different from that in weak coupling calculus. 

To see this, firstly, let us recall the 
prepotentials in weak coupling region. In this case, the WDVV equations 
were satisfied even at perturbative level. \cite{MMM,MMM2,IY2,BMMM} 
Namely, the perturbative 
prepotentials can be obtained by solving WDVV equations at 
perturbative level as was explicitly 
shown by Braden {\em et al.} \cite{BMMM} in the case of SU(4) gauge theory. 

\begin{center}
\subsection{Perturbative dual prepotential of SU($r+1$) gauge theory}
\end{center}

In the case of strongly coupled theory, on the other hand, the dual perturbative 
prepotentials themselves do not satisfy (\ref{duawdvv}), thus 
in this case the dual WDVV equations do not hold at perturbative level. 

To check this, let us recall the dual prepotential of SU($r+1$) gauge theory 
obtained from study of period integrals. \cite{DS,DP} 
According to the result, the perturbative part of dual 
prepotential $\F_{D,\mbox{\scriptsize per}}$ can be represented by a 
single function $f$ 
	\beq
	\F_{D,\mbox{\scriptsize per}}=\sum_{i=1}^{r} f(a_{D_i})
	.\lab{pp}
	\eeq
As the argument of $f$ is single, the matrices $(\F_{D_i})$ in 
(\ref{duawdvv}) become singular, e.g., the only non-zero entry 
of $(\F_{D_1})$ is 
$(\F_{D_1})_{11} =\pa^3 \F_{D,\mbox{\scriptsize per}}/\pa a_{D_1}^3$. 
This indicates that we can not determine $\F_{D,\mbox{\scriptsize per}}$ 
from (\ref{duawdvv}) even if we follow the method of 
Braden {\em et al.}. \cite{BMMM} 

\begin{center}
\subsection{Non-perturbative dual prepotential of SU(4) gauge theory}
\end{center}

Then, what happens when the non-perturbative part is introduced? 
In this case, we add the non-perturbative part 
$\F_{D,\mbox{\scriptsize non}}$ to $\F_{D,\mbox{\scriptsize per}}$ 
and consider 
	\beq
	\F_{D}=\F_{D,\mbox{\scriptsize per}}+
	\F_{D,\mbox{\scriptsize non}}
 	,\lab{no}
	\eeq
where 
	\beq
	\F_{D,\mbox{\scriptsize non}}=\sum_{k=1}^{\infty}
	\F_{D,k}\l^{k}
	\lab{333}
	\eeq
and $\l^{-1}\equiv \l_{\mbox{\scriptsize SU($r+1$)}}$ is 
the dynamically generated mass 
scale of SU($r+1$) gauge theory. In (\ref{333}), 
the coefficients $\F_{D,k}$ are functions in dual variables $a_{D_i}$. 

As it is not easy to study the general case of $r$, we restrict $r=3$ case 
in the following discussion. In this case, substituting (\ref{no}) into 
(\ref{duawdvv}), we can obtain 
nothing from the coefficient of $\l^0$, but we can find from 
the coefficient of $\l^{1}$ 
	\beq
	f^{\,'''}(a_{D_1})f^{\,'''}(a_{D_2})f^{\,'''}(a_{D_3})
	\pa_1 \pa_2 \pa_3 \F_{D,1}=0
	\lab{1eq}
	\eeq
and from that of $\l^{2}$ 
	\beqa
	& &f^{\,'''}(a_{D_1})f^{\,'''}(a_{D_2}) \left( \pa_2 \pa_{3}^2 
	\F_{D,1} \cdot\pa_1 \pa_{3}^2 \F_{D,1}-\pa_{3}^3 \F_{D,1}\cdot
	\pa_1 \pa_2 \pa_3 \F_{D,1}\right)\nm\\	
	& &-f^{\,'''}(a_{D_1})f^{\,'''}(a_{D_3}) \left( \pa_{2}^3 
	\F_{D,1} \cdot \pa_1 \pa_2 \pa_{3}\F_{D,1}-\pa_{2}^2\pa_3 \F_{D,1}
	\cdot\pa_1 \pa_{2}^2\F_{D,1}\right)\nm\\	
	& &+f^{\,'''}(a_{D_2})f^{\,'''}(a_{D_3}) \left( \pa_{1}^2 \pa_{3} 
	\F_{D,1} \cdot\pa_{1}^2 \pa_{2}\F_{D,1}-
	\pa_{1}^3 \F_{D,1}\cdot\pa_1 \pa_2 \pa_3 \F_{D,1}\right)\nm\\
	& &-f^{\,'''}(a_{D_1})f^{\,'''}(a_{D_2})f^{\,'''}(a_{D_3})
	\pa_1 \pa_2 \pa_3 \F_{D,2}=0
	,\lab{2eq}
	\eeqa
where $f^{\,'''}(a_{D_i})=d^3 f(a_{D_i})/da_{D_i}^3$ and 
$\pa_i \equiv \pa /\pa a_{D_i}$. It is interesting to notice that in the 
weak coupling calculus of SU(4) gauge theory one-instanton prepotential 
satisfies a complicated equation \cite{O2} while in the present case the equation 
for $\F_{D,1}$ is very simple. 

To calculate $\F_{D,k}$ explicitly, the perturbative prepotential must be fixed, 
although it is not available directly from (\ref{duawdvv}) in contrast with 
the weak coupling study. 
For this reason, we must provide it as the input data. Actually, 
the perturbative part is known to be calculated as \cite{DS,DP} 
	\beq
	f(a_{D_i})=a_{D_i}^2 \ln \frac{a_{D_i}}
	{\l_{\mbox{\scriptsize SU(4)}}}
	\lab{3d}
	,\eeq
where we have ignored the overall numerical factor and 
the normalization of $\l_{\mbox{\scriptsize SU(4)}}$ in (\ref{3d}). 
In this case, the third-order derivatives of (\ref{3d}) do not vanish, so 
the general solution to (\ref{1eq}) is easily calculated to give 
	\beq
	\F_{D,1}=f_1 (a_{D_2} ,a_{D_3})+f_2 (a_{D_1},a_{D_3})+
	f_3 (a_{D_1},a_{D_2})
	,\lab{3eq}
	\eeq	
where $f_i$ are arbitrary functions. 

For $\F_{D,2}$, on the other hand, from (\ref{2eq}) and (\ref{3eq}), we get 
	\beqa
	\F_{D,2}&=&\frac{1}{2}\int (a_{D_1}\pa_{1}^2 \pa_3 f_2 \cdot 
	\pa_{1}^2 \pa_2 f_3 +a_{D_2}\pa_1 \pa_{2}^2 f_3 \cdot 
	\pa_{2}^2 \pa_3 f_1 +a_{D_3}\pa_2 \pa_{3}^2 f_1 \cdot \pa_1 
	\pa_{3}^2 f_2 )da_{D_1}da_{D_2}da_{D_3}\nm\\
	& &+
	g_1 (a_{D_2} ,a_{D_3})+g_2 (a_{D_1},a_{D_3})+
	g_3 (a_{D_1},a_{D_2})
	,\lab{388}
	\eeqa
where we have again used arbitrary functions $g_i$.  

Here, let us notice that the scaling relation for $\F_{D,k}$ is given by 
	\beq
	\sum_{i=1}^3 a_{D_i}\frac{\pa \F_{D,k}}{\pa a_{D_i}}=
	(2+k)\F_{D,k}
	\lab{scal}
	\eeq
which follows from dimensional analysis. 
The scaling relation \cite{Mat,STY,EY,HS,KO} for dual 
prepotential \cite{DP,IY3,O1} was a basic tool of the study of strong 
coupling expansion presented by D'Hoker and Phong. \cite{DP} 

Of course, though there are various functions satisfying (\ref{scal}), 
we can easily see that all monomials of 
degree $3$ for (\ref{3eq}) are also solutions to (\ref{scal}) by 
following to the method presented in weak coupling study of WDVV 
equations, \cite{O2} thus we get 
	\beq
	\F_{D,1}=\sum_{i=1}^{3}s_i a_{D_i}^3 +c_1 a_{D_1}^2 a_{D_2}+c_2 
	a_{D_1}^2 a_{D_3} +c_3 a_{D_2}^2 a_{D_3} +c_4 a_{D_1} a_{D_2}^2 
	+c_5 a_{D_1}a_{D_3}^2 +c_6 a_{D_2}a_{D_3}^2 
 	\lab{449} 
	,\eeq
where $s_i$ and $c_i$ are integration constants. 
The function form of (\ref{449}) is consistent to the result of SU(4) 
gauge theory obtained by D'Hoker and Phong, \cite{DP} but note that the 
integration constants should be determined by other approaches. A priori, 
we do not know explicit values for them in view of (\ref{duawdvv}). 

In a similar manner, we can determine $\F_{D,2}$ from (\ref{388}) as 
	\beqa
	\F_{D,2}&=&\sum_{i=1}^3 t_i a_{D_i}^4 +c_1 c_2 a_{D_1}^2 a_{D_2} a_{D_3}+c_3 c_4 
	a_{D_1} a_{D_2}^2 a_{D_3}+c_5 c_6 a_{D_1} a_{D_2} a_{D_3}^2 \nm\\
	& &+\left\{ a_{D_1}a_{D_2}^3 ,\ a_{D_1}^2 a_{D_2}^2 ,\ 
	a_{D_1}^3 a_{D_2},\ a_{D_1}a_{D_3}^3 ,\ a_{D_1}^2 a_{D_3}^2 ,\ 
	a_{D_1}^3 a_{D_3} ,\ a_{D_2}a_{D_3}^3 ,\ a_{D_2}^2 a_{D_3}^2 ,\ 
	a_{D_2}^3 a_{D_3} \right\}
	,\eeqa
where $t_i$ are integration constants, 
the braces mean any linear combination of the elements and 
we have assumed that $g_i$ consist of polynomials.

{\bf Remark:} {\em In general, it is known from explicit 
examples \cite{KLT,IY3,IY6} that $\F_{D,k}$ are represented by 
polynomials in dual periods. }

\begin{center}
\section{Summary}
\end{center}

In this paper, we have considered the consequence of electro-magnetic 
duality transformation for the WDVV equations (\ref{wdvv}) 
and derived the dual 
WDVV equations (\ref{duawdvv}) satisfied by dual prepotentials. The 
dual WDVV equations are turned out to have the same form with the original 
WDVV equations, but the perturbative part of dual prepotential 
do not satisfy dual WDVV equations. However, we have derived the 
non-perturbative dual prepotential in pure SU(4) gauge theory as an 
example by appropriately introducing perturbative part and following to 
the method to get solutions developed in weak coupling calculus. \cite{O2} 
In fact, we have found that there is 
the non-perturbative prepotential in strong coupling region which is 
consistent to the result of D'Hoker and Phong. \cite{DP} From this 
result, it is important to notice that we can study both weak and 
strong coupling prepotentials in the standpoint of WDVV 
equations. 

As for an another direction to study the strong coupling region, 
it may be interesting to try to develop the 
topological string theoretic interpretation \cite{IXY} in 
strong coupling region and to search a connection to (\ref{duawdvv}). 
More detailed study should be expected in the future. 

\begin{center}

\end{center}


\begin{thebibliography}{99}

\bibitem{SW1}
N. Seiberg and E. Witten, 
Nucl. Phys. B {\bf 431}, 484 (1994). 

\bibitem{SW2}
N. Seiberg and E. Witten, 
Nucl. Phys. B {\bf 435}, 129 (1994).

\bibitem{Sei}
N. Seiberg, 
Phys. Lett. B {\bf 206}, 75 (1988). 

\bibitem{KLT}
A. Klemm, W. Lerche, and S. Theisen, 
Int. J. Mod. Phys. A {\bf 11}, 1929 (1996). 

\bibitem{FP}
D. Finnell and P. Pouliot, 
Nucl. Phys. B {\bf 453}, 225 (1995).

\bibitem{DKM}
N. Dorey, V. V. Khoze, and M. P. Mattis,  
Phys. Lett. B {\bf 388}, 324 (1996). 

\bibitem{DKM2}
N. Dorey, V. V. Khoze, and M. P. Mattis, 
Phys. Rev. D {\bf 54}, 2921 (1996).  

\bibitem{GKMMM}
A. Gorsky, I. Krichever, A. Marshakov, A. Mironov, and A. Morozov, 
Phys. Lett. B {\bf 355}, 466 (1995). 

\bibitem{Nek}
N. Nekrasov, 
Nucl. Phys. B {\bf 531}, 323 (1998). 

\bibitem{Wit}
E. Witten, 
Nucl. Phys. B {\bf 340}, 281 (1990). 

\bibitem{DW}
R. Dijkgraaf and E. Witten, 
Nucl. Phys. B {\bf 342}, 486 (1990). 

\bibitem{DVV}
R. Dijkgraaf, E. Verlinde, and H. Verlinde, 
Nucl. Phys. B {\bf 352}, 59 (1991). 

\bibitem{Dub}
B. Dubrovin, 
``Geometry of 2D topological field theories,'' 
hep-th/9407018.

\bibitem{MMM}
A. Marshakov, A. Mironov, and A. Morozov, 
Phys. Lett. B {\bf 389}, 43 (1996). 

\bibitem{MMM2}
A. Marshakov, A. Mironov, and A. Morozov, 
``More evidence for the WDVV equations in $N=2$ SUSY 
Yang-Mills theories,'' hep-th/9701123. 

\bibitem{MMM3}
A. Marshakov, A. Mironov, and A. Morozov, 
Mod. Phys. Lett. A {\bf 12}, 773 (1997). 

\bibitem{IY2}
K. Ito and S.-K. Yang, 
Phys. Lett. B {\bf 433}, 56 (1998). 

\bibitem{BMMM}
H. W. Braden, A. Marshakov, A. Mironov, and A. Morozov, 
Phys. Lett. B {\bf 448}, 195 (1999). 

\bibitem{O2}
Y. Ohta, 
``One-instanton prepotentials from WDVV equations in $N=2$ 
supersymmetric SU(4) Yang-Mills theory,'' 
hep-th/9904125. 

\bibitem{DS}
M. Douglas and S. Shenker, 
Nucl. Phys. B {\bf 448}, 271 (1995). 

\bibitem{DP}
E. D'Hoker and D. H. Phong, 
Phys. Lett. B {\bf 397}, 94 (1997). 

\bibitem{KLYT}
A. Klemm, W. Lerche, S. Yankielowicz and S. Theisen, 
Phys. Lett. B {\bf 344}, 169 (1995). 

\bibitem{APS}
P. C. Argyres and A. E. Faraggi, 
Phys. Rev. Lett. {\bf 74}, 3931 (1995). 

\bibitem{HO}
A. Hanany and Y. Oz, 
Nucl. Phys. B {\bf 452}, 283 (1995). 

\bibitem{DS1}
U. H. Danielsson and B. Sundborg, 
Phys. Lett. B {\bf 358}, 273 (1995). 

\bibitem{Han}
A. Hanany, 
Nucl. Phys. B {\bf 466}, 85 (1996). 

\bibitem{BL}
A. Brandhuber and K. Landsteiner, 
Phys. Lett. B {\bf 358}, 73 (1995). 

\bibitem{Mat}
M. Matone, 
Phys. Lett. B {\bf 357}, 342 (1995).  

\bibitem{STY}
J. Sonnenschein, S. Theisen, and S. Yankielowicz, 
Phys. Lett. B {\bf 367}, 145 (1996). 

\bibitem{EY}
T. Eguchi and S.-K. Yang, 
Mod. Phys. Lett. A {\bf 11}, 131 (1996). 

\bibitem{HS}
P. S. Howe and P. C. West, 
Nucl. Phys. B {\bf 486}, 425 (1997). 

\bibitem{KO}
H. Kanno and Y. Ohta, 
Nucl. Phys. B {\bf 530}, 73 (1998).

\bibitem{IY3}
K. Ito and S.-K. Yang, 
``Picard-Fuchs equations and prepotentials in $N=2$ 
supersymmetric QCD,'' hep-th/9603073. 

\bibitem{O1}
Y. Ohta, 
J. Math. Phys. {\bf 40}, 1891 (1999).

\bibitem{IY6}
K. Ito and S.-K. Yang, 
Phys. Lett. B {\bf 366}, 165 (1996). 

\bibitem{IXY}
K. Ito, C.-S.Xiong, and S.-K. Yang, 
Phys. Lett. B {\bf 441}, 155 (1998). 

\end{thebibliography}
\end{document}